\documentclass[reprint,pra,groupedaddress,showpacs,twocolumn]{revtex4}
\usepackage{units}
\usepackage{amsmath,mathtools}
\usepackage{amssymb}
\usepackage{graphicx}
\usepackage{bm,microtype,color}

\newcommand{\be}{\begin{equation}}
\newcommand{\ee}{\end{equation}}
\newcommand{\bea}{\begin{eqnarray}}
\newcommand{\eea}{\end{eqnarray}}
\newcommand{\im}{\text{Im}}
\newcommand{\re}{\text{Re}}
\newcommand{\cc}[1]{{\color{black}{#1}}}

\begin{document}
\title{Rotation-induced Mode Coupling in Open Wavelength-scale Microcavities}

\author{Li Ge}
\email{li.ge@csi.cuny.edu}
\affiliation{\textls[-20]{Department of Engineering Science and Physics, College of Staten Island, CUNY, Staten Island, NY 10314, USA}}
\affiliation{\textls[-20]{The Graduate Center, CUNY, New York, NY 10016, USA}}
\author{Raktim Sarma}
\affiliation{Department of Applied Physics, Yale University, New Haven, CT 06520-8482, USA}
\author{Hui Cao}
\email{hui.cao@yale.edu}
\affiliation{Department of Applied Physics, Yale University, New Haven, CT 06520-8482, USA}

\date{\today}
\begin{abstract}
We study the interplay between rotation and openness for mode coupling in wavelength-scale microcavities. In cavities deformed from a circular disk, the decay rates of a quasi-degenerate pair of resonances may cross or anti-cross with increasing rotation speed. The standing-wave resonances evolve to traveling-wave resonances at high rotation speed, however, both the clockwise (CW) and counter-clockwise (CCW) traveling-wave resonances can  have a lower cavity decay rate, in contrary to the intuitive expectation from rotation-dependent effective index. With increasing rotation speed, a phase locking between the CW and CCW wave components in a resonance takes place. These phenomena result from the rotation-induced mode coupling, which is strongly influenced by the openness of the microcavity. The possibility of a non-monotonic Sagnac effect is also discussed.
\end{abstract}
\maketitle

\section{Introduction}
Eigenmodes are fundamental in understanding both quantum and wave phenomena. When the system is perturbed, the eigenmodes of the original, unperturbed system become coupled. In optics, for example, the coupling can be introduced by matter-mediated interaction in cavity quantum electrodynamics \cite{cQED1,cQED2}, by nonlinearity in multimode lasers \cite{Science08}, and by linear scattering from a local defect or a gradual boundary deformation in optical waveguides \cite{waveguide} and microcavities \cite{Wiersig_PRA06,Wiersig_PRL06,BD1,BD2}. In addition, rotation causes a minute change of the refractive index \cite{Sarma_JOSAB2012}, which leads to the mixing of standing-wave resonances in optical microcavities \cite{Harayama_PRA06,Harayama_OpEx07}. The well-known Sagnac effect \cite{Sagnac,Scully,Aronowitz,Ciminelli,Terrel}, i.e. the rotation-induced frequency splitting, has also been reported in microcavities \cite{Sarma_JOSAB2012,Harayama_PRA06,Harayama_OpEx07,Scheuer2}. Although most microcavities have open boundaries, the openness or coupling to the environment has not been considered as a key factor that can dictate the behaviors of rotating cavities; it has only been studied as a quantity that can be influenced by the rotation \cite{Sarma_JOSAB2012,Scheuer2}. In microcavities much larger than the wavelength of the resonances, the effect of the openness is weak and the cavity can be treated as a closed system. This treatment is not sufficient for wavelength-scale microcavities \cite{Zhang,Song_PRL10,Redding_PRL12}, which are valuable for integrated photonics circuits, among others, because of their small footprints and mode volumes.

In this report we show that the openness of wavelength-scale microcavities can have strong influence on rotation-induced phenomena, including the Sagnac effect. We first show analytically that it slightly enhances the Sagnac effect in circular microcavities. Its effect is much stronger in asymmetric resonant cavities (ARCs) \cite{Nockel,Gmachl,wiersig_prl08}, which can lead to different scenarios of mode coupling, including crossing of the cavity decay rates and a non-monotonic frequency splitting with rotation. These behaviors are analyzed using a coupled-mode theory, and two key quantities are identified, i.e. the phase of the coupling constant between the quasi-degenerate resonances at rest, and the phase of the difference in their complex resonant frequencies. With closed boundaries both phases vanish, thus their non-zero values result from the openness of the cavity. Our analysis also reveals a passive phase locking between the clockwise (CW) and counterclockwise (CCW) traveling wave components in a resonance at high rotation speed.

Below we focus on transverse magnetic (TM) resonances in two-dimensional (2D) microcavities without loss of generality. Their electric field is in the cavity plane, and their magnetic field, represented by $\psi(\vec{r})$, is perpendicular to the cavity plane. To the leading order of the rotation speed $\Omega$, the resonances $\psi$ and their frequencies $k$ of an {\it open} cavity are determined by the modified Helmholtz equation \cite{Sarma_JOSAB2012}
\begin{gather}
\left[\nabla^2 + \epsilon(\vec{r})k^2 + 2ik\frac{\Omega}{c}\frac{\partial}{\partial \theta}\right]\psi(\vec{r}) = 0,\label{eq:0}
\\
\epsilon(\vec{r}) =
\begin{cases}
n^2, & r<\rho(\theta) \\
1, & r>\rho(\theta)
\end{cases}\nonumber
\end{gather}
where $\rho(\theta)$ is the boundary of the microcavity in the polar coordinates and the origin is at the rotation center. $c$ is the speed of light in vacuum, and $n$ is the refractive index inside the cavity. We have assumed that the rotation axis is perpendicular to the cavity plane and $\Omega>0$ indicates CCW rotation. Eq.~(\ref{eq:0}) is a generalization for closed cavity modes discussed in Ref.~\cite{Harayama_PRA06}, and the openness of the cavity makes the resonant frequencies $k$ complex, with an negative imaginary part that reflects the cavity decay rate, i.e. $\kappa = -2\im[k]>0$.

\section{circular microdisk cavities}

We start with a circular dielectrtic disk of radius $R$. The angular momentum $m$ is a conserved quantity. A pair of CW ($\psi\propto e^{- i|m|\theta}$) and CCW resonances ($\psi\propto e^{i|m|\theta}$) are degenerate when the cavity is stationary, with the same complex resonant frequency $k_{0}$. The angular momentum is still conserved at a nonzero rotation speed $\Omega$,
i.e. Eq.~(\ref{eq:0}) can be solved by imposing the following ansatz
\be
\psi(\vec{r})=\begin{cases}
J_m(\bar{k}_mr)e^{im\theta}, & r<R \\
H^+_m(\tilde{k}_mr)e^{im\theta}, & r>R
\end{cases}
\ee
where $\bar{k}_m \equiv \sqrt{n^2k^2-2km{\Omega}/{c}} =  n(k-{m\Omega}/n^2c) + O(|\Omega/c|^2)$, $\tilde{k}_m \equiv \sqrt{k^2-2km{\Omega}/{c}} =  k-{m\Omega}/c + O(|\Omega/c|^2)$, and $J_m,H_m^+$ are the Bessel function and Hankel function of the first kind. $k$ is determined by
\be
\bar{k}_m\frac{J'_m(\bar{k}_mR)}{J_m(\bar{k}_mR)} = \tilde{k}_m\frac{{H^+_m}'(\tilde{k}_mR)}{H^+_m(\tilde{k}_mR)},\label{eq:BD}
\ee
which is required by the continuity of $\psi(\vec{r})$ and its radial derivative at $r=R$.
Eq.~(\ref{eq:0}) was studied numerically in Ref.~\cite{Sarma_JOSAB2012} using a finite-different-time-domain (FDTD) method adapted to the rotating frame. The results show that the aforementioned double-degenerate resonances of frequency $k_0$ at rest split at infinitesimal $\Omega$, and the differences in both real and imaginary parts of their complex frequencies  increase linearly with $\Omega$, with an enhanced Sagnac effect (i.e. for the real part) compared with {\it closed} microcavities \cite{Harayama_PRA06}.

\begin{figure}[t]
\centering
\includegraphics[width=\linewidth]{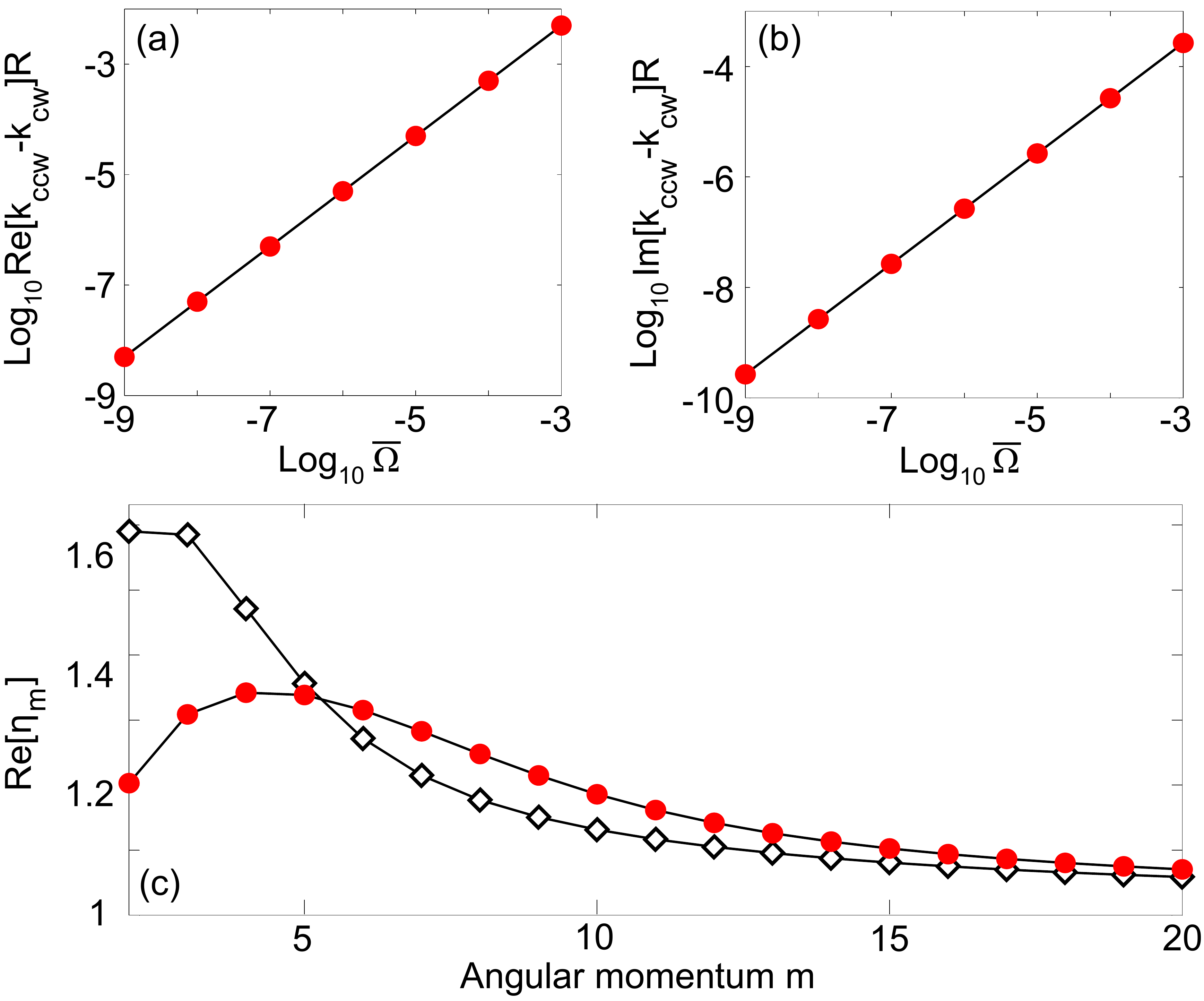}
\caption{(Color online) Splitting of the real (a) and imaginary (b) parts of the complex frequencies for a pair of $|m|=8$ resonances at $k_0R\simeq5.3923 - 0.0114i$ in a circular microdisk of refractive index $n=2$, plotted as a function of the normalized rotation speed. Symbols show numerical solutions of Eq.~(\ref{eq:BD}) and solid lines give the perturbation result (\ref{eq:eta}) with $\eta_8\simeq1.2478 + 0.0665i$. (c) Enhancement of the Sagnac effect in open circular microdisk cavities given by $\re[\eta_m]$. Connected dots and diamonds are for $n=2,3$, respectively.} \label{fig:disk}
\end{figure}

Below we confirm these results analytically by expanding Eq.~(\ref{eq:BD}) to the leading order of the dimensionless rotation speed $\overline\Omega\equiv R\Omega/c$, which reveals that the slightly enhanced Sagnac effect in an open cavity depends on the angular momentum.
We derive from Eq.~(\ref{eq:BD}) that
\begin{gather}
kR = k_0R + \frac{m\overline\Omega}{n^2}\eta_m + O(|\overline\Omega|^2),\label{eq:eta}\\
\eta_m = \frac{m^2}{k_0^2R^2} - \left[\frac{{H^+_m}'(k_0R)}{H^+_m(k_0R)}\right]^2.
\end{gather}
Note that both $k$ and $\eta_m$ are complex due to the openness of the cavity.
It can be shown that for whispering-gallery modes $\re[\eta_m]$ is larger than 1 and it approaches this lower bound as $|m|\rightarrow\infty$ [see Fig.~\ref{fig:disk}(c)]. Thus the Sagnac effect in an open cavity is slightly enhanced from its values in a closed cavity, i.e.
\be
\re[k_{ccw}-k_{cw}] \approx \frac{2|m|\Omega}{n^2c}\re[\eta_m],\label{eq:Sagnac}
\ee
and the enhancement factor $\re[\eta_m]$ is stronger in wavelength-scale microcavities where $|m|$ is small.
We note that this does not imply that the Sagnac effect itself is stronger in wavelength-scale cavities, since the dominant dependence still comes from the linear size of the cavity, reflected by the factor of $|m|$ in Eq.~(\ref{eq:Sagnac}). For a small refractive index inside the cavity (e.g. $n=2$), the $|m|$-dependence of $\re[\eta_m]$ is non-monotonic, and $\re[\eta_m]$ reaches a local maximum at a certain $m$ [Fig.~\ref{fig:disk}(c)].

The approximation (\ref{eq:eta}) agrees well with the numerical solutions of Eq.~(\ref{eq:BD}). One example is given in Fig.~\ref{fig:disk}(a) and (b), in which $n=2$, $k_0R\simeq5.3923 - 0.0114i$ and we found $\eta_8\simeq1.2478 + 0.0665i$, indicating that the splitting of the imaginary parts of the complex resonances is about 20 times smaller than that of the real parts. But since $\re[k_0]/|\im[k_0]|\sim50$, the relative change of the splitting of the imaginary parts is larger compared with the real parts, as found numerically in Ref.~\cite{Sarma_JOSAB2012}.
We also note that the mixing of the CW and CCW waves of the same $|m|$ found in Ref.~\cite{Sarma_JOSAB2012} is not caused by rotation but rather by the way of excitation in the FDTD method, as we have shown that each resonance contains only CW or CCW wave of a single $m$.

In the discussion above we have assumed that the rotation axis is at the center of the microdisk cavity. Eq.~(\ref{eq:0}) still holds when the disk center is away from the rotation axis (which is the origin of the polar coordinates by definition), and the circular microdisk cavity becomes an ARC since now $\rho(\theta)\neq const$. We will study ARCs in general in the next section.

\section{Asymmetric resonant cavities}

There are several ways to find the resonances in a rotating ARC. In addition to the modified Finite-Difference-Time-Domain (FDTD) simulation \cite{Sarma_JOSAB2012}, a perturbative approach can be employed in any numerical methods that incorporates an outgoing boundary condition, such as
\be
\psi(\vec{r}) = \sum_m \gamma_m H^+_m(\tilde{k}_mr)e^{im\theta},\quad r>\rho(\theta). \label{eq:outgoing}
\ee
One example is the Finite-Difference-Frequency-Domain method used in Ref.~\cite{Science08}, where the cavity is put inside a  circular computational domain. For any realistic value of the rotation speed, $\overline\Omega\ll1$ and the gradient term on the left hand side of Eq.~(\ref{eq:0}) only leads to a small shift of the resonant frequencies. Thus a perturbative root search can be implemented by first calculating the resonant frequencies of the stationary disk ($k_0$), approximating $k$ by $k_0$ in the gradient term, calculating the resulting $k$, inserting it back to the gradient term, and repeat the process until $k$ converges.

Here we employ a nonperturbative approach, the modified scattering matrix method proposed in Ref.~\cite{rotCav_PRL}. Beside the consideration of numerical efficiency, one motivation is to capture the cavity shape exactly. It was recently found that even a minute perturbation on the scale of one thousandth of the wavelength can cause a drastic variation of the emission pattern in wavelength-scale microcavities \cite{BD1}. Finite difference or finite element method unavoidably introduces a small deviation when approximating the smooth cavity boundary by discrete grids, while the scattering matrix method utilizes the analytical form of the cavity boundary and is free of spatial grids.

The scattering matrix method applies to a concave cavity with a uniform refractive index and a smooth boundary deviation $\delta\rho(\theta)$ from a circle satisfying the Rayleigh criterion $|\delta\rho(\theta)|\ll R$.
In this approach the wave function of a resonance inside the cavity is decomposed in the angular momentum basis, i.e.
\be
\psi(\vec{r})=\sum_m [\alpha_mH^+_m(\bar{k}_mr) + \beta_mH^-_m(\bar{k}_mr)]e^{im\theta},\label{eq:expansion}
\ee
where $H^-$ are the Hankel functions of the second kind. Outside the cavity the outgoing condition (\ref{eq:outgoing}) is used.
Compared with the formulation for non-rotating cavities \cite{Narimanov_PRL1999,Smatrix}, the difference lies in the $m$-dependent frequencies $\bar{k}_m$ and $\tilde{k}_m$ defined previously.

Defining the vectors $|\alpha\rangle$, $|\beta\rangle$, $|\gamma\rangle$ from the coefficients in Eqs.~(\ref{eq:outgoing}) and (\ref{eq:expansion}), the regularity of $\psi(\vec{r})$ at the origin is satisfied by requiring $|\alpha\rangle=|\beta\rangle$. The continuity conditions of $\psi(\vec{r})$ and its radial derivative at $\rho(\theta)$ can be put into the following matrix form
\begin{gather}
\mathcal{\overline{H}}^+|\alpha\rangle + \mathcal{\overline{H}}^-|\beta\rangle = \mathcal{\tilde{H}}^+|\gamma\rangle, \label{eq:BC1}\\
\mathcal{\overline{D}}^+|\alpha\rangle + \mathcal{\overline{D}}^-|\beta\rangle = \mathcal{\widetilde{D}}^+|\gamma\rangle,\label{eq:BC2}
\end{gather}
in which
\begin{align}
[\mathcal{\overline{H}}^\pm]_{lm} = \int_0^{2\pi} H^\pm_m(\bar{k}_m\rho(\theta)) e^{i(m-l)\theta} d\theta, \\
[\mathcal{\overline{D}}^\pm]_{lm} = \int_0^{2\pi} \bar{k}_m{H^\pm}'_m(\bar{k}_m\rho(\theta)) e^{i(m-l)\theta} d\theta,
\end{align}
and $\mathcal{\tilde{H}}^+, \, \mathcal{\widetilde{D}}^+$ are defined similarly with $\tilde{k}_m$ in place of $\bar{k}_m$. By eliminating $|\gamma\rangle$ from Eqs.~(\ref{eq:BC1}) and (\ref{eq:BC2}), a matrix equation can be found in the form $\mathcal{S}(k)|\alpha\rangle = |\beta\rangle$. By taking into account the constraint $|\alpha\rangle=|\beta\rangle$ mentioned above, we solve $\mathcal{S}(k)|\alpha\rangle = |\alpha\rangle$ \cite{Narimanov_PRL1999,Smatrix} to find the resonances $k$. Below we exemplify the effect of the openness on the Sagnac effect and chiral symmetry breaking in a wavelength-scale lima\c{c}on cavity using this method.

\subsection{Chiral symmetry breaking and emission pattern asymmetry}
\label{sec:chiral}
It was found that spontaneous breaking of chiral symmetry occurs in wavelength-scale microcavities \cite{Zhang,Song_PRL10, Redding_PRL12}: CW and CCW waves in a resonance follow symmetric but distinct orbits. This is due to the wave effect of light, which cannot be treated as rays traveling in straight lines and undergoing specular reflections at the cavity boundary. We found that such orbits evolve into resonances dominated by CW or CCW waves at large $\Omega$, with only small variations of their intensity patterns inside and outside the cavity.

\begin{figure}[t]
\centering
\includegraphics[width=\linewidth]{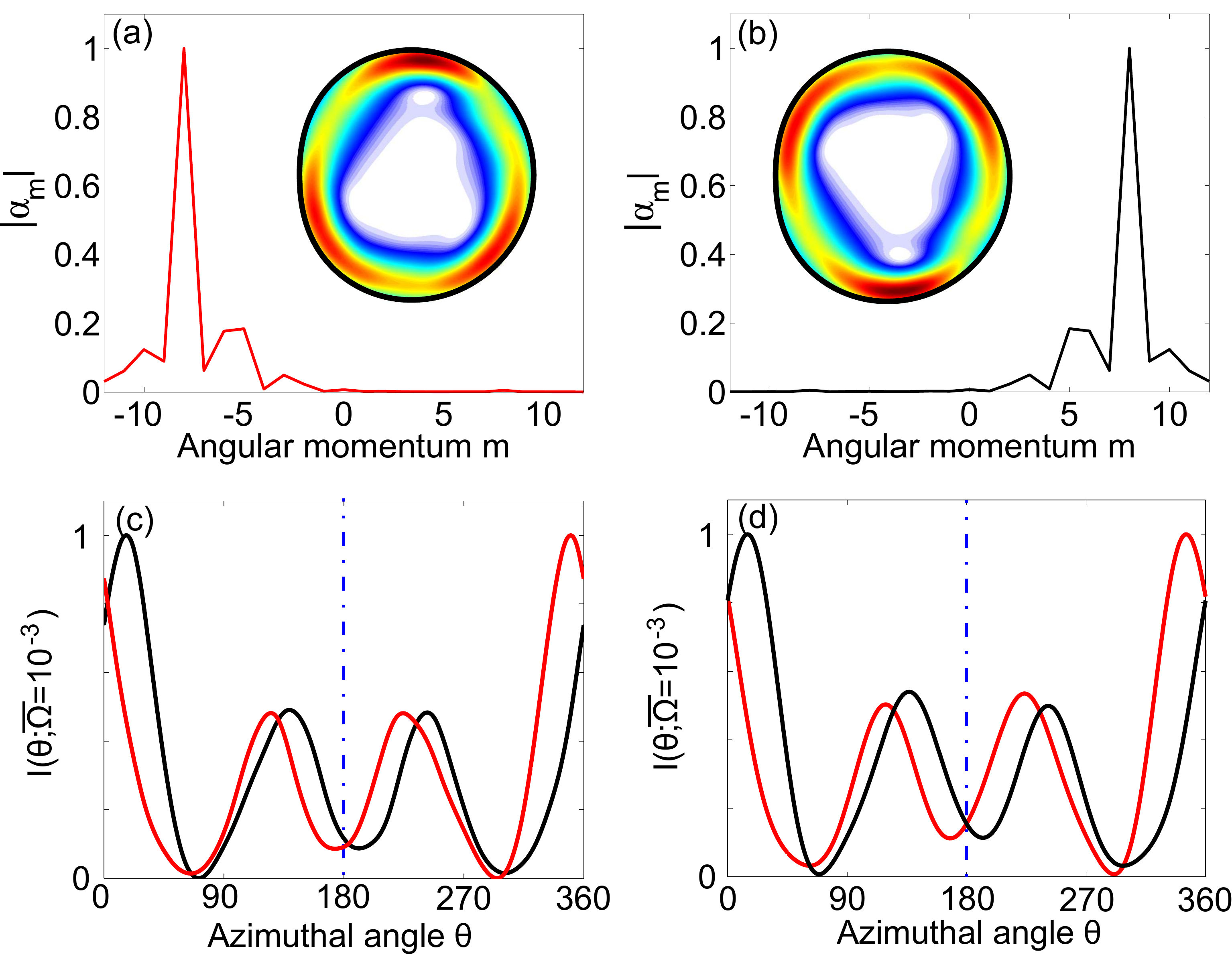}
\caption{(Color online) Normalized expansion coefficients $\alpha_m$ defined in Eq.~(\ref{eq:expansion}) for a pair of CW-dominated (a) and CCW-dominated (b) resonances at $\overline\Omega=10^{-3}$, which have a dominant angular momentum $m=-8, \, 8$, respectively. The cavity is a lima\c{c}on with $\epsilon=0.41$ and $n=2$, shifted along $\theta=180^\circ$ by $R\epsilon$ such that $\rho(0)=\rho(180^\circ)$. The insets show the intracavity intensity patterns of these resonances, which are almost chiral symmetric. They cannot be distinguished by eye from those of the CW and CCW waves in the corresponding standing-wave resonances at rest (not shown). (c) The angular dependence of their external field intensities at $r=3R$. Note that they are not mirror images of each other about the symmetry axis of the cavity (dash-dotted line), especially near $\theta=0(360^\circ),180^\circ$. (d) Same as (c) but for the CW waves (red) and CCW waves (black) in the corresponding standing-wave resonances at rest. They are mirror images of each other about the symmetry axis of the cavity. } \label{fig:chiral}
\end{figure}

One example is shown in Fig.~\ref{fig:chiral} using a lima\c{c}on cavity, the boundary of which is given by $\rho(\theta) = R(1+\epsilon\cos\theta)$. The deformation from a circle (due to a finite $\epsilon$) breaks the degeneracy of the resonances at rest, and each standing-wave resonance now has multiple angular momenta, with a dominant pair ($m,-m$) if the deformation is small.
Because the lima\c{c}on is symmetric about the horizontal axis ($\theta=0,\,180^\circ$), the wave functions of these standing-wave resonances are either even or odd about this axis, which we denote by $\psi^+$ and $\psi^-$. They appear as quasi-degenerate pairs, with each pair having the same dominant angular momenta ($m,-m$). We will refer to the pair with $|m|=8$ in Fig.~\ref{fig:chiral} as Pair 1.

Fig.~\ref{fig:chiral}(c) and (d) shows the similarity of the external field intensity $I(\theta;\overline\Omega)$ for CW and CCW waves at $\bar\Omega=0,10^{-3}$ and $r=3R$. However, they are not exactly the same. This can be seen from the chiral symmetry at  $\bar\Omega=0$, i.e. $I_{cw}(\theta;\overline\Omega=0) =  I_{ccw}(-\theta;\overline\Omega=0)$, and the lack of it between $I_{cw}$ and $I_{ccw}$ at $\overline\Omega=10^{-3}$.
The latter is true for the intensity patterns inside the cavity as well, and in general  $\psi_{ccw}(r,\theta;\overline\Omega)\neq\psi_{cw}(r,-\theta;\overline\Omega)$, even though $\psi_{ccw}(r,\theta;\overline\Omega)=\psi_{cw}(r,-\theta;-\overline\Omega)$ as can be seen from Eq.~(\ref{eq:0}).

We note that $I(\theta;\overline\Omega)$ has a weak $r$-dependent even in the asymptotic region \cc{$r\gg R^2 / \lambda$}. This is because the argument of the Hankel functions in the expansion (\ref{eq:outgoing}) outside the cavity is $m$-dependent, thus the factor $\exp(i\tilde{k}_mr)/\sqrt{\tilde{k}_mr}$ in the asymptotic form of the Hankel functions is not a common factor for all angular momenta, in contrast to the stationary case. We have considered a rotation speed much slower than $c/r$ such that Eq.~(\ref{eq:0}) is valid. For a faster rotation the higher-order terms $O(\overline\Omega^2)$ neglected in Eq.~(\ref{eq:0}) can be significant in the far field, which may cause an additional $r$-dependency of the far-field emission pattern.

\begin{figure}[b]
\centering
\includegraphics[width=\linewidth]{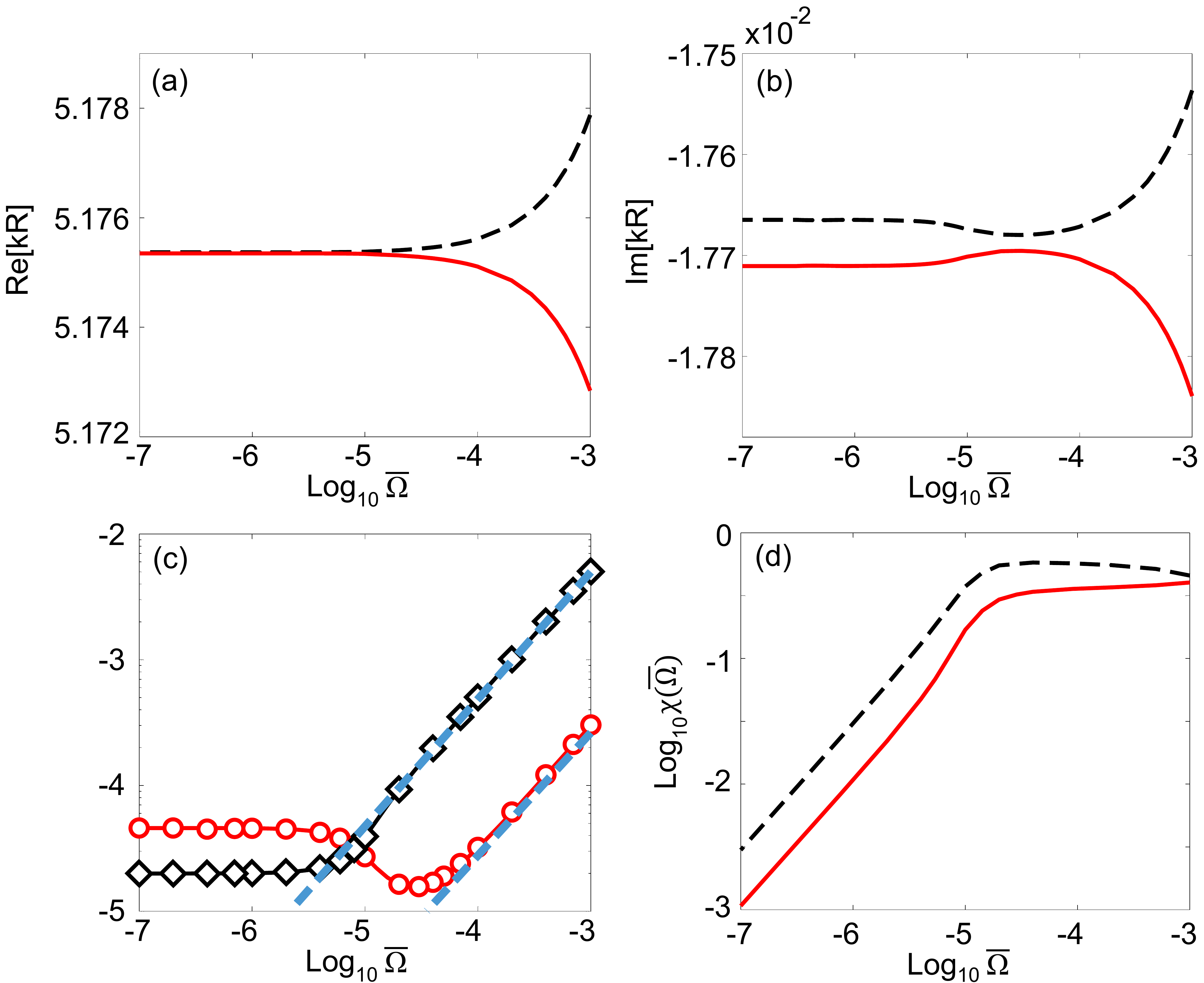}
\caption{(Color online) Mode coupling of Pair 1 shown in Fig.~\ref{fig:chiral}. (a, b) Real and imaginary parts of the complex frequencies of the resonances resonances that evolve into the CW- (red solid) and CCW-dominated (black dashed) ones.
The angular dependence of their external field intensities is shown in (d). The symbols in (c) show their splitting $(k_{ccw}-k_ {cw})R$ (real part: diamonds; imaginary part: circles) in the logarithmic scale ($\log_{10}$), where the solid lines are given by the coupled-mode theory (\ref{eq:Deltak}) with $g = 4.99 + 0.30i$ and the dashed lines show the complex frequency splitting of the $|m|=8$ resonances in a circular cavity of the same radius.
} \label{fig:antiX}
\end{figure}

As shown in Fig.~\ref{fig:antiX}(a) and (c), the splitting of the resonant frequencies in an open ARC displays a threshold, similar to the Sagnac effect in closed microcavities \cite{Harayama_PRA06}. However, the asymmetry $\chi(\overline\Omega)$ of the emission pattern, which can be characterized by
\be
\chi(\overline\Omega) = \frac{\int_{0}^\pi \;I(\theta;\overline\Omega)d\theta}{\int_{\pi}^{2\pi} I(\theta;\overline\Omega)d\theta}-1,
\ee
does not have a threshold at low $\overline\Omega$; it displays an almost linear dependence on $\overline\Omega$ until the wave function becomes dominated by either CW or CCW waves [Fig.~\ref{fig:antiX}(d)], similar to the finding in larger cavities with $|m|\sim100$ \cite{rotCav_PRL}. This was explained using a coupled-mode theory \cite{rotCav_PRL}, which we employ in the next section to study the $\overline\Omega$-dependence of the complex resonant frequencies, especially the non-monotonic behaviors of their imaginary parts shown in Fig.~\ref{fig:antiX}(b) and (c).

\subsection{Rotation-induced mode coupling}

CCW rotation ($\Omega>0$) increases the effective index inside and outside the cavity for CW waves ($m<0$) \cite{Sarma_JOSAB2012}:
\be
n_\text{eff}
\equiv \left\{
\begin{matrix*}[l]
\dfrac{\bar{k}_m}{k} = \sqrt{n^2-2m\dfrac{\overline\Omega}{kR}},\quad &r<\rho(\theta), \\[0.7em]
\dfrac{\tilde{k}_m}{k} = \sqrt{1-2m\dfrac{\overline\Omega}{kR}}, \quad &r>\rho(\theta).
\end{matrix*}
\right.\label{eq:neff}
\ee
Therefore, we expect the resonant frequencies of CW-dominated resonances to reduce as a function of $\Omega$.
Meanwhile, we expect their cavity decay rates (given by $-2\text{Im}[k]$) to increase, since the index contrast at the cavity boundary is reduced. The situation is reversed for CCW waves. Thus for a circular microdisk cavity both $\re[k_{ccw}-k_{cw}]$ and $\im[k_{ccw}-k_{cw}]$ are positive when the cavity undergoes a CCW rotation and they increase with $\Omega$. These intuitive expectations are verified numerically in Ref. \cite{Sarma_JOSAB2012} and analytically in Fig.~\ref{fig:disk}(a), (b).

\begin{figure}[t]
\centering
\includegraphics[width=\linewidth]{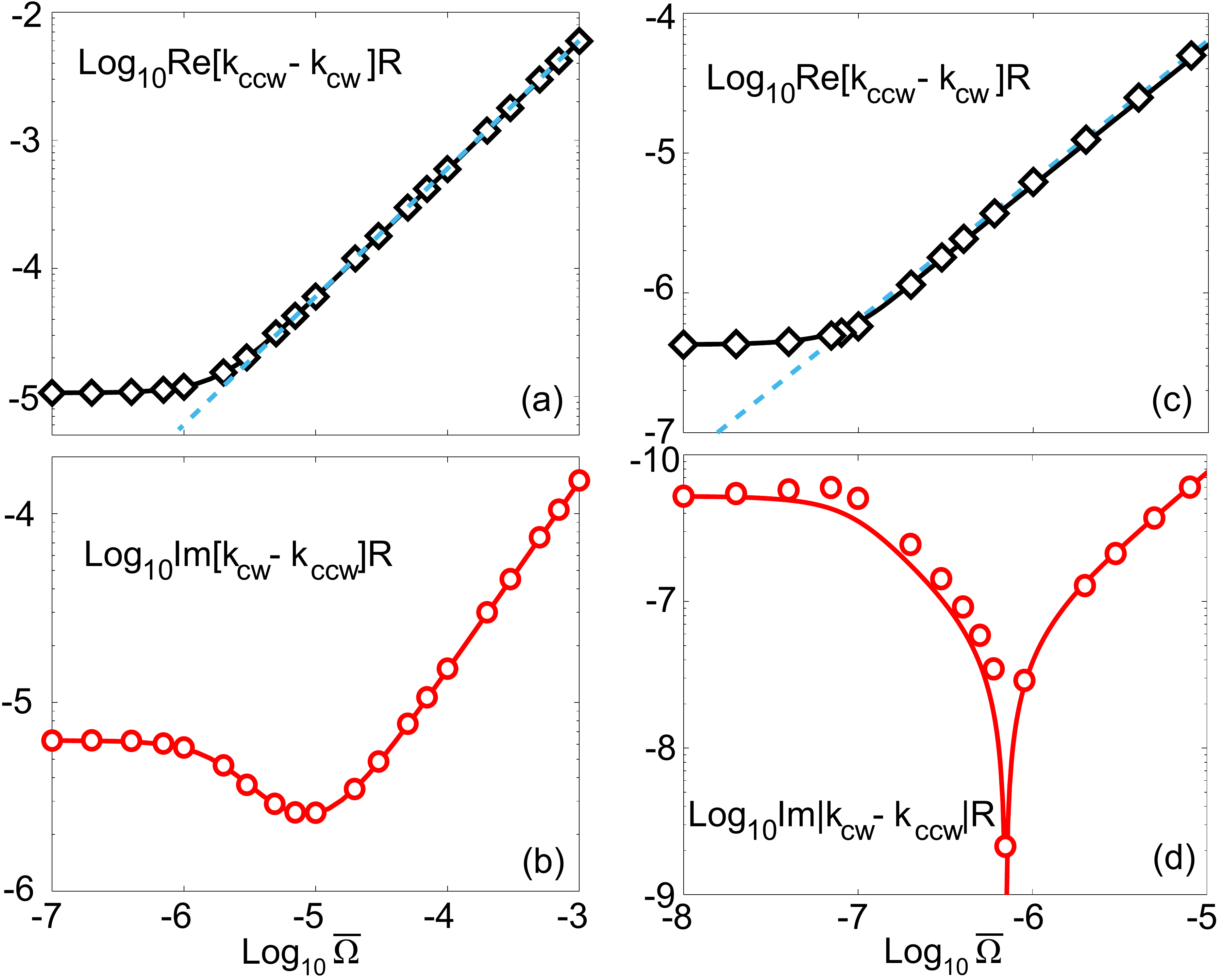}
\caption{(Color online) The same as Fig.~\ref{fig:antiX}(c) but for Pair 2 with a dominant $|m|=10$ (a,\,b) and Pair 3 with a dominant $|m|=11$ (c,\,d). Solid lines are given by the coupled-mode theory (\ref{eq:Deltak}) with $g = 5.92 - 0.15i,\,5.90-0.071i$, respectively. Dashed lines in (a) and (c) show the splitting of real part of the frequencies for the corresponding resonances in a circular cavity of the same radius.
In (b) the cavity decay rates anti-cross and $|\im[k_{ccw}]|>|\im[k_{cw}]|$. In (d) they cross and $|\im[k_{ccw}]|>|\im[k_{cw}]|$ for $\overline\Omega\gtrsim10^{-6.2}$.} \label{fig:antiX_invert}
\end{figure}

For the quasi-degenerate resonances Pair 1 of the lima\c{c}on cavity shown in Fig.~\ref{fig:chiral}, these expectations also hold at large $\Omega$ [see Fig.~\ref{fig:antiX}(a) and (b)]. It is surprising, however, that $\im[k_{ccw}],\im[k_{cw}]$ undergo an avoided crossing at an intermediate $\Omega$. The same behavior is observed for the next pair with a dominant angular momentum $|m|=9$ (not shown).
More surprisingly, we found that for the resonances with a dominant $|m|=10$ (Pair 2), the CW-dominated mode has a {\it lower} cavity decay rate at large $\Omega$ [Fig.~\ref{fig:antiX_invert}(a)], and the intuitive prediction based on the index contrast fails.
The same holds for the resonances with a dominant $|m|=11$ (Pair 3) but now with a crossing of the cavity decay rates [Fig.~\ref{fig:antiX_invert}(b)].

To understand these behaviors, we resort to the coupled-mode theory described in Ref.~\cite{rotCav_PRL}. It is similar to that developed in Refs.~\cite{Harayama_PRA06,Harayama_OpEx07,Sunada_PRA08}, but it is adopted to open cavities, taking into account the non-vanishing phases of the coupling constant $g$ between the quasi-degenerate resonances $\psi^+,\psi^-$ at rest and the difference of their complex resonant frequencies $k^+_0,k^-_0$, which we will show to be the key quantities that determine the different behaviors of the cavity decay rates mentioned above.

As the cavity rotates, the resonances become CW- or CCW-dominated, which can be viewed as the result of the coupling of the corresponding standing-wave resonances $\psi^+$ and $\psi^-$ at rest, 
i.e. $\psi(\Omega) \approx a^+(\Omega)\psi^+ + a^-(\Omega)\psi^-$. Eq.~(\ref{eq:0}) can then be rewritten as a coupled-mode equation
\be
\left(
\begin{array}{c c}
k^2-{k_{0}^+}^2 & \frac{2ik\Omega}{cn^2}G_{\!\scriptscriptstyle +-} \\
\frac{2ik\Omega}{cn^2}G_{\!\scriptscriptstyle -+} & k^2-{k_{0}^-}^2
\end{array}\right)
\begin{pmatrix}
a^+ \\ a^-
\end{pmatrix}
=0,\label{eq:modecoupling}
\ee
where $G_{\!\scriptscriptstyle +-} \equiv \int_\text{cavity} \psi^+ {\partial_\theta\psi^-}\,d\vec{r}$ and $G_{\!\scriptscriptstyle -+}$ is defined similarly. We note that $G_{\!\scriptscriptstyle ++}$ and $G_{\!\scriptscriptstyle --}$, which would have appeared on the diagonal of the coupling matrix in Eq.~(\ref{eq:modecoupling}), vanish because their integrands are odd functions with respect to the horizontal axis. Likewise, $\int_\text{cavity} \psi^+\psi^-\,d\vec{r}$ vanishes even though resonances of an open cavity are not orthogonal or biorthogonal in general. We have used the normalization $\int_\text{cav} (\psi^{\pm})^2\,d\vec{r}=1$. The difference of the two resonances $k_{cw}, k_{ccw}$ at rotation speed $\Omega$ is given by
\be
\Delta k (\Omega) = \left[(\Delta k_0)^2 + \left(\frac{g}{c}\Omega\right)^2\right]^{\frac{1}{2}}, \label{eq:Deltak}
\ee
where $\Delta k_0=k_0^--k_0^+$ and $g$ is the dimensionless coupling constant defined by
$g\equiv2\sqrt{-G_{\!\scriptscriptstyle -+}G_{\!\scriptscriptstyle +-}}/n^2$.

Equation (\ref{eq:Deltak}) shows that the frequency splitting, both its real and imaginary parts, is very small for $\Omega$ smaller than the critical value $\Omega_c\equiv c|\Delta k_0/g|$, below which the leading $\Omega$-dependence is quadratic; $\Delta k(\Omega)$ is reduced by a factor of $\Omega/2\Omega_c$ when compared with a circular microdisk, where $\Delta k_0=0$ and the leading $\Omega$-dependence is linear. Far beyond $\Omega_c$, $\Delta k(\Omega)$ approaches its asymptote $g\Omega/c$, and its real part gives the Sagnac frequency splitting, which is similar to the value of the corresponding resonances in a circular microdisk of the same radius [see the dashed lines in Figs.~\ref{fig:antiX}(c) and \ref{fig:antiX_invert}(a),(c)].
We also note that the sum of $k_{cw},k_{ccw}$ is given by the same expression (\ref{eq:Deltak}) but with $\Delta k_0$ replaced by $k_0^+ + k_0^-$. Since $|g|\Omega/c\ll |k_{cw}|,|k_{ccw}|$ for any realistic rotation speed, the sum (and the average) of $k_{cw},k_{ccw}$ only has a leading $O(\overline\Omega^2)$ dependence even beyond $\overline\Omega_c$, which is weaker than the rotation dependence of their splitting for $\overline\Omega>\overline\Omega_c$. This explains why the real and imaginary parts of the complex frequencies shown in Fig.~\ref{fig:antiX}(a) and (b) look symmetric about their average.

The coupling constant $g$ is approximately real and positive in a cavity slightly deformed from a circular disk. This can be seen from its definition, and more specifically, the relation that $G_{\!\scriptscriptstyle -+}\approx-G_{\!\scriptscriptstyle +-}$.
The minute phase of $g$ is due to the openness of the cavity, and it determines whether the CW- or CCW-dominated resonance has a lower cavity decay rate for $\Omega\gg\Omega_c$. This can be seen by substituting $(a^+,a^-)$ in Eq.~(\ref{eq:modecoupling}) by $(1,-i)$ for a CW-dominated resonance and $(1,i)$ for a CCW-dominated resonance, leading to
\begin{gather}
k_{cw}(\Omega) \rightarrow \frac{k_0^+ + k_0^- - g\frac{\Omega}{c}}{2},\label{eq:kcw}\\
k_{ccw}(\Omega) \rightarrow \frac{k_0^+ + k_0^- + g\frac{\Omega}{c}}{2}\label{eq:kccw}.
\end{gather}
Therefore, the CW-dominated resonances have a lower frequency and higher cavity decay rate \textit{only if} $g$ is the in the first quadrant of the complex plane. This is the case for Pair 1 shown in Fig.~\ref{fig:antiX}, and a good fit is given by $g=4.99+0.30i$. $g$ is in the fourth quadrant for both Pair 2 and 3 (fitted with $g=5.92-0.15i,\, 5.90-0.07i$ in Fig.~\ref{fig:antiX_invert}), and as a result the CW-dominated resonances have a lower frequency {\it and} a lower cavity decay rate for $\Omega\gg\Omega_c$.

In view of these findings, the failure of the predictions based on the effective index (\ref{eq:neff}) is understandable since it does not consider the interference between $\psi^+$ and $\psi^-$, which changes as a function of the rotation speed. We note that the coupled-mode theory (\ref{eq:modecoupling}) does not apply to circular cavities, because the angular momentum is still a good quantum number as mentioned previously, which can only be achieved by a {\it fixed} combination of $\psi^+\propto\cos(m\theta)$ and $\psi^-\propto\sin(m\theta)$.
\cc{We also note that the value of $g$ calculated by integrating the wave functions obtained from the scattering matrix method agrees well with the value extracted from fitting the complex resonance splitting in large cavities \cite{rotCav_PRL}. In the wavelength-scale microcavities studied here, additional wave effects (such as multimode coupling \cite{BD2}) are present and the two values of $g$ only agree qualitatively; the calculated value of $g$ is $4.09+0.03i,5.00-0.18i,5.34-0.15i$ for Pair 1, 2, and 3, respectively. Nevertheless, it is important to note that the calculated value and the fitting value of $g$ for the same pair of resonances are in the same quadrant of the complex plane and close to the real axis.}

The minute phase of $g$, together with the phase of $\Delta k_0$, also determines whether the cavity decay rates of a pair of resonances cross each other. This can be understood by inspecting Eq.~(\ref{eq:Deltak}): crossing of the decay rates take place when the sum in the square root, denoted by $\Sigma(\Omega)$,
becomes a real positive number at some value of $\Omega$.
A necessary condition is that $g$ and $\Delta k_0$ are in neighboring quadrants in the complex plane, which guarantees that $\Sigma(\Omega)$ can become real. Note that this criterion does not depend on whether the CW-dominated resonance originates from the parity-odd resonance $\psi^-$ or the parity-even resonance $\psi^+$, or in other words, whether $\Delta k(\Omega=0)$ is given by $\Delta k_0$ or $-\Delta k_0$. For Pair 1, $\Delta k_0R \simeq (1.99+4.59i)\times10^{-5}$ and $g=4.99+0.27i$ are both the first quadrant; for Pair 2, $\Delta k_0R \simeq (1.07-0.58i)\times10^{-5}$ and $g=5.92-0.15i$ are both in the fourth quadrant. Therefore, for these two pairs their respective decay rates do not cross each other.

To find the sufficient condition for the crossing, we note again that $g$ is almost real in a cavity slightly deformed from a circular cavity.
The sufficient condition for the cavity decay rates to cross is completed by the requirement that
the acute angle formed between $\Delta k_0$ and the imaginary axis, denoted by $\angle (\Delta k_0,\pm i)$, is larger than $|\text{Arg}[g]|$, where $\text{Arg}$ denotes the principle value of the phase in $(-\pi,\pi]$.
For Pair 3, $\Delta k_0R \simeq (4.25+5.23i)\times10^{-7}$ is in the first quadrant while $g=5.90-0.071i$ is in the fourth quadrant, satisfying the necessary condition. In addition, $\angle (\Delta k_0,\pm i)=0.68>|\text{Arg}[g]|=0.012$, which completes the sufficient condition and leads to the crossing of the cavity decay rates.

\begin{figure}[t]
\centering
\includegraphics[width=\linewidth]{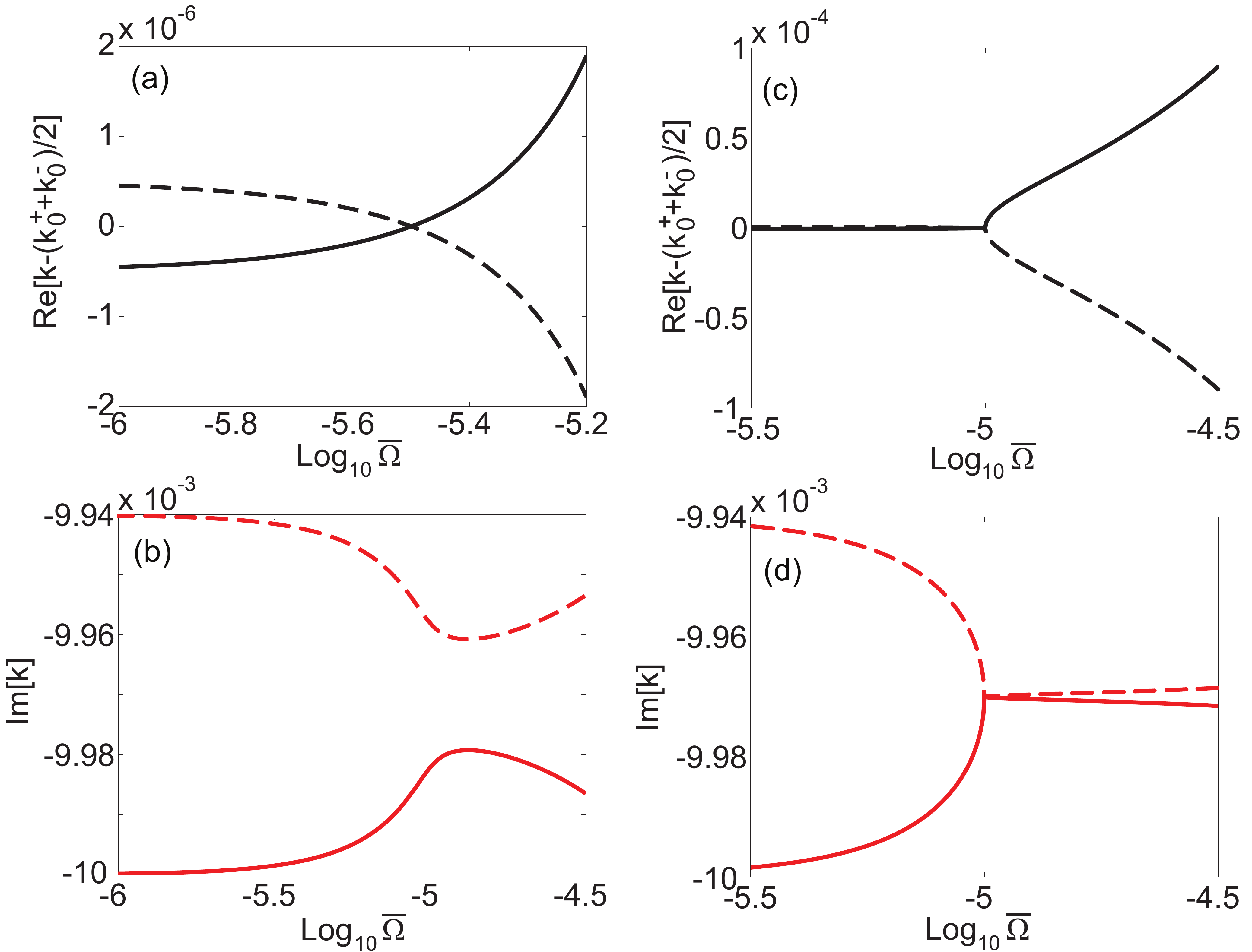}
\caption{(Color online) (a,b) Crossing of the real parts of a pair of quasi-degenerate resonances, constructed using Eq.~(\ref{eq:modecoupling}) and with $g=6-1i$, $\Delta k_0R=(0.1+6i)\times10^{-5}$, $k_0^-R=5-0.01i$. (c,d) A pair of quasi-degenerate resonances reach an exceptional point at $\overline\Omega=\overline\Omega_c=10^{-5}$, constructed using Eq.~(\ref{eq:modecoupling}). The parameters are the same as in (a,b) except for $g=6-0.1i$.}
\label{fig:EP}
\end{figure}

We note that crossing of the real part of $\Delta k(\Omega)$ is also possible in principle [Fig.~\ref{fig:EP}(a)], which means that the Sagnac frequency splitting is no longer a monotonic function of the rotation speed. It occurs when $\Sigma(\Omega)$ becomes negative at some value of $\Omega$. It still requires the same necessary condition that $g$ and $\Delta k_0$ are in neighboring quadrants in the complex plane, which guarantees that $\Sigma(\Omega)$ can become real. In addition, it requires that $\angle (\Delta k_0,\pm i)<|\text{Arg}[g]|$. An even more dramatic scenario can take place in principle, if $\Sigma(\Omega)$ becomes zero at some value of $\Omega$. It requires that $\Delta k_0$ and $g$ are $\pm\pi/2$ out of phase with each other, and when this holds, \cc{the two resonances reach an exceptional point \cite{EP} at $\Omega=\Omega_c$, with identical complex resonant frequencies and wave functions.} If the phase of $g$ is really small, then an approximate bifurcation happens for $\re[k]$ and an approximate inverse bifurcation happens for $\im[k]$ [Fig.~\ref{fig:EP}(c),(d)], due to a phase singularity (a jump by $\pi$) of $\Sigma(\Omega)$. These two scenarios discussed here and shown in Fig.~\ref{fig:EP} require that $\Delta k_0$ to be essentially imaginary, which may be realized by fine-tuning the cavity shape.

\subsection{Phase locking between CW and CCW waves}

\begin{figure}[t]
\centering
\includegraphics[width=\linewidth]{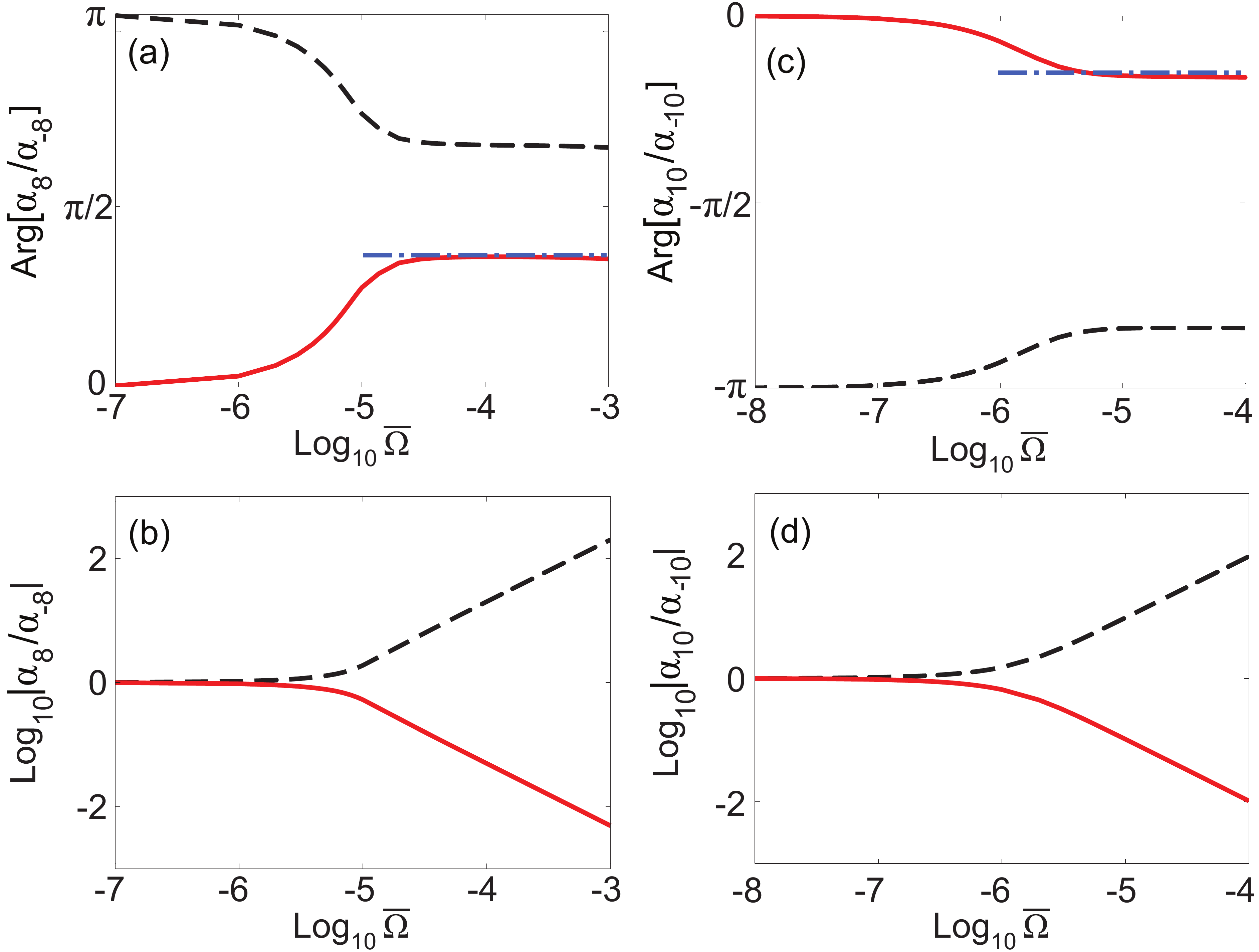}
\caption{(Color online) Relative phase (a) and amplitude (b) of the dominant angular components $m=\pm8$ in Pair 1 as a function of $\Omega$. Red solid and black dashed lines represent the resonances that evolve into CW- or CCW-dominated ones, respectively. (c) and (d) show the same for the dominant angular components $m=\pm10$ in Pair 2. Dash-dotted lines in (a) and (c) are given by $\Delta\varphi=1.11,-0.471$, respectively, which are given by Eq.~(\ref{eq:phaseLock}) from the coupled-mode theory.} \label{fig:phaseLock}
\end{figure}

Finally, we report a passive phase locking between CW and CCW waves in a resonance as the rotation speed increases.
As Fig.~\ref{fig:phaseLock}(a) and (c) shows, the relative phase between $\alpha_{|m|}$ (CCW) and $\alpha_{-|m|}$ (CW) at rest is either 0 or $\pm\pi$, which gives the parity-even and parity-odd resonances. As the cavity rotates, this relative phase gradually approaches a locked value $\Delta \varphi$ for $\Omega>\Omega_c$.
$\Delta \varphi$ is in $[0,\pi/2]$ for the CW-dominated resonance in Pair 1, and it is in $[-\pi/2,0]$ for the CW-dominated resonance in Pair 2. This difference seems to be related to whether the CW- or CCW-dominated resonances have a higher cavity decay rates, or equivalently, whether the coupling constant $g$ between $\psi^+$ and $\psi^-$ is in the first or fourth quadrant.

To confirm this relationship, we again resort to the coupled-mode equation (\ref{eq:modecoupling}), which gives the mixing ratio $\xi(\Omega) \equiv {a^-}/{a^+}$ for a pair of quasi-degenerate resonances \cite{rotCav_PRL}
\begin{align}
\xi(\Omega)^2
\approx \frac{D \pm\sqrt{D^2+\left( {2g^2}/{c^2} \right) \Xi\Omega^2}}{D \mp\sqrt{D^2+ \left( {2g^2}/{c^2} \right) \Xi\Omega^2}},
\label{eq:mixing}
\end{align}
where $D\equiv {k_0^-}^2-{k_0^+}^2$ and $\Xi\equiv {k_0^-}^2+{k_0^+}^2$. It is straightforward to show that the second term in the radicand dominates when $\Omega>\Omega_c$, and in this limit we can further approximate
\be
\xi(\Omega)^2 \approx \frac{\frac{\pm c\Delta k_0}{g\Omega}-1}{\frac{\pm c\Delta k_0}{g\Omega}+1} \label{eq:xi2b}
\ee
 by also taking into account that $D/\sqrt{2\Xi}\approx\Delta k_0$. \cc{For a CW-dominated resonance with a dominant angular momentum $-|m|$ and a locked phase $\Delta\varphi$, its wave function can be approximated by $\psi(\Omega) \approx \zeta\exp(i\Delta \phi)\exp(i|m|\theta) + \exp(-i|m|\theta)$ with a real $\zeta\equiv|\alpha_{|m|}/\alpha_{-|m|}|\ll1$. or in other words,
\be
\xi(\Omega)^2
\rightarrow\frac{2\zeta\exp(i\Delta \varphi)-1}{2\zeta\exp(i\Delta \varphi)+1}\label{eq:xi2}
\ee
as $\Omega$ becomes much larger than $\Omega_c$. By comparing with Eq.~(\ref{eq:xi2b}), we immediately find $\zeta \approx {\Omega_c}/{\Omega}$ and the locked phase is given by
\be
\Delta\varphi \approx \text{Arg}\left[\frac{\pm\Delta k_0}{g}\right].\label{eq:phaseLock}
\ee}

It is clear from Eq.~(\ref{eq:phaseLock}) that $\Delta\varphi$ is not determined by the phase of $g$ alone but also by that of $\Delta k_0$. The latter is more influential in cavities slightly deformed from a circular disk, where $g$ is almost real and positive as mentioned previously. \cc{The ``$\pm$" signs in Eq.~(\ref{eq:phaseLock}) come from the two possibilities that either $\psi^+$ or $\psi^-$ evolves into a CW-dominated resonance. This uncertainty can change $\Delta\varphi$ by $\pi$}, but it does not mix the two different scenarios found in Fig.~\ref{fig:phaseLock}(a) and (c), i.e. whether $\Delta\varphi\in[0,\pi/2]$ or $[-\pi/2,0]$. We find that the positive sign in Eq.~(\ref{eq:phaseLock}) corresponds to the locked phase for the CW-dominated resonance in Pair 1 and 2, which gives $\Delta\varphi=1.11,-0.471$, respectively. They agree well with the numerical results shown in Fig.~\ref{fig:phaseLock}(a) and (c).

The locked phase in the CCW-dominated resonance can be found similarly, which gives $\Delta\varphi \approx \text{Arg}[\mp g/\Delta k_0]$, and the sum of the two locked phases in these resonances is approximately $\pm\pi$.
The latter feature can be easily identified in Fig.~\ref{fig:phaseLock}(a) and (c).

\section{Conclusion}

In summary, we have shown both analytically with the coupled-mode theory and numerically with a scattering matrix method that the openness of wavelength-scale microcavities has a strong effect on rotation-induced mode coupling. Openness results in non-vanishing phases of the coupling constant $g$ and the complex frequency splitting $\Delta k_0$ of the quasi-degenerate resonances at rest. These two quantities together dictate the rotation dependence of the decay rates and the resonant frequencies. The decay rates of the quasi-degenerate resonances may cross or anti-cross with increasing rotation speed, and unlike the circular microcavities, both the CW- or CCW-dominated resonances of asymmetric resonant cavities can have a lower cavity decay rate, depending on the phase of $g$. The well-known Sagnac effect, i.e. the linear increase of resonant frequency splitting with the rotation speed, may be altered by mode coupling and exhibit a non-monotonic behavior. Finally, the relative phase of the CW and CCW wave components in a resonance is locked at high rotation speed as a result of mode coupling. These unusual behaviors of mode coupling result from the interplay between openness and rotation in wavelength-scale microcavities.

\section*{Acknowledgment}

We thank Takahisa Harayama and Jan Wiersig for helpful discussions. L.G. acknowledges PSC-CUNY 45 Research Award. R.S. and H.C. acknowledges NSF support under Grant No. ECCS-1128542.

\bibliographystyle{apsrev}

\end{document}